\documentclass[a4paper,11pt]{article}
\pdfoutput=1
\usepackage{jheppub}

\usepackage{epsf}
\usepackage{epsfig}
\usepackage{subfigure}
\usepackage{mathtools}
\usepackage{hhline}
\usepackage{float}
\usepackage{multirow}
\usepackage{nicefrac}
\usepackage{epstopdf}
\usepackage{slashed}
\usepackage{xcolor}
\usepackage{url}

\newcommand{\be}{\begin{eqnarray}}
\newcommand{\ee}{\end{eqnarray}}

\newcommand{\ba}{\begin{array}}
\newcommand{\ea}{\end{array}}
\newcommand{\bee}{\begin{equation}\ba{c}}
\newcommand{\eee}{\ea\end{equation}}

\newcommand{\bi}{\begin{itemize}}
\newcommand{\ei}{\end{itemize}}

\toccontinuoustrue

\title{New decay modes of heavy Higgs bosons in a two Higgs doublet model with vectorlike leptons}
\author{Radovan Derm\'i\v{s}ek$^{1,2}$,}
\author{Enrico Lunghi$^1$}
\author{and Seodong Shin$^1$}
\affiliation{
$^1$Physics Department, Indiana University, Bloomington, IN 47405, USA \\
$^2$Department of Physics and Astronomy and Center for Theoretical Physics, Seoul National University, Seoul 151-747, Korea \\
}
\emailAdd{dermisek@indiana.edu} 
\emailAdd{elunghi@indiana.edu} 
\emailAdd{shinseod@indiana.edu}

\abstract{In models with extended Higgs sector  and additional matter fields, the decay modes of  heavy Higgs bosons can be dominated by cascade decays through the new fermions rendering present search strategies ineffective. We investigate new decay topologies of heavy neutral Higgses in two Higgs doublet model with vectorlike leptons. We also discuss constraints from existing searches and discovery prospects. Among the most interesting  signatures are monojet, mono $Z$, mono Higgs, and $Z$ and Higgs bosons produced with a pair of charged leptons.

\newpage
 
}
\preprint{
\begin{minipage}{3cm}
\small
\flushright
IU-HET-609
\end{minipage}} 

\begin{document}

\maketitle

\section{Introduction}
\label{sec:introduction}

In models with extended Higgs sector and additional matter fields, the decay patterns of heavy Higgs bosons can be dramatically altered   limiting the potential of present search strategies or rendering them nugatory. However, new search strategies can be designed that could lead to a simultaneous discovery of heavy Higgs bosons and matter particles, and that are more potent than separate searches for such particles. Such a situation occurs already in a two Higgs doublet model with vectorlike leptons.
  
\begin{figure}
\begin{center}
\includegraphics[width=0.8\linewidth]{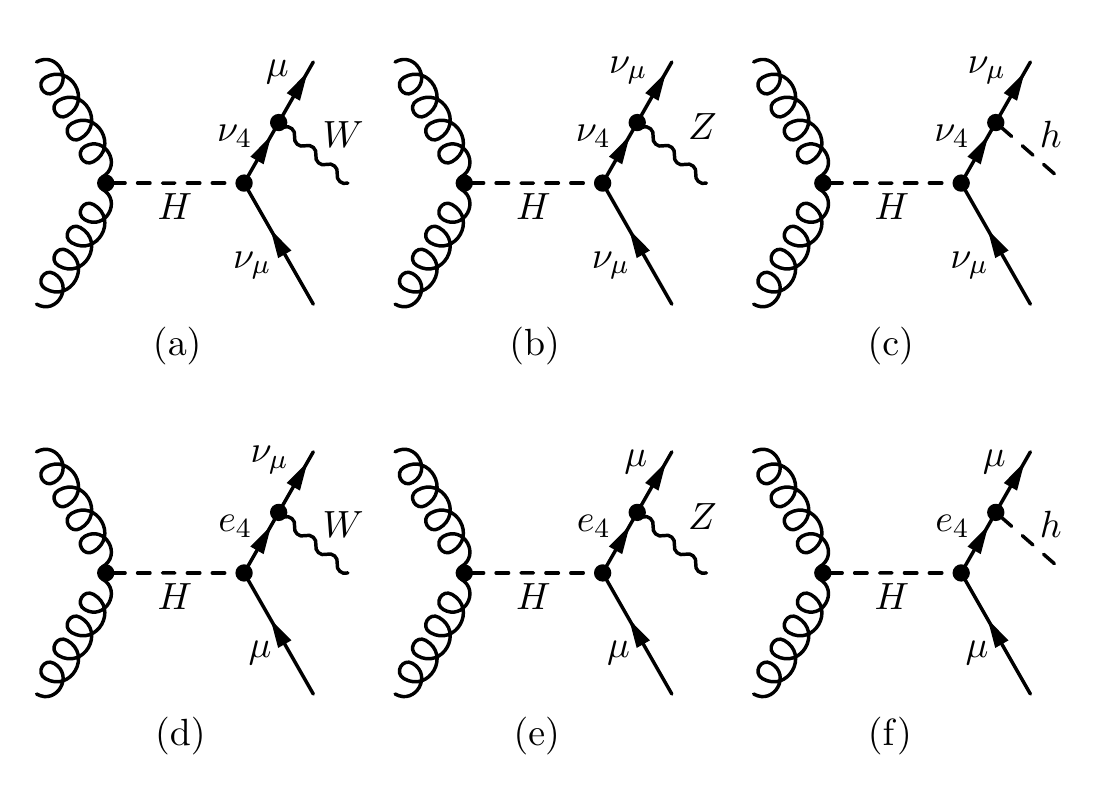}
\caption{New decay topologies of the heavy Higgs boson.
\label{fig:topologies}
}
\end{center}
\end{figure}

We consider an extension of the two Higgs doublet model type-II by vectorlike pairs of new leptons introduced in ref.~\cite{Dermisek:2015oja}. In this model, the new leptons mix only with one family of standard model (SM)  leptons and we will use the second family as an example. As a result of the  mixing of new vectorlike leptons with leptons in the SM  the  flavor changing couplings of  $W$, $Z$ and Higgs bosons between heavy and light leptons are generated. These couplings allow new decay modes for heavy CP even (or CP odd) Higgs boson: $H \to \nu_4 \nu_\mu$ and $H \to e_4 \mu$, where $e_4$ and $\nu_4$ are the  lightest new charged and neutral leptons. These decay modes can be very large when the mass of  the heavy Higgs boson is below the $t \bar t$ threshold and the light Higgs boson ($h$) is SM-like so that $H \to ZZ,\; WW$ are suppressed or not present. In this case, flavor changing decays $H \to \nu_4 \nu_\mu$ or $H \to e_4 \mu$ compete only with $H\to b \bar b$  and for sufficiently heavy $H$ also with $H \to hh$. 
Subsequent decay modes of $e_4$ and $\nu_4$: $e_4 \to W  \nu_\mu$, $e_4 \to Z \mu$,  $e_4 \to h \mu$ and $\nu_4 \to W \mu$, $\nu_4 \to Z \nu_\mu$, $\nu_4 \to h \nu_\mu$ lead to the following 6 decay chains of the heavy Higgs boson:
\begin{align}
H &\; \to \; \nu_4 \nu_\mu \; \to \; W\mu \nu_\mu, \; Z\nu_\mu \nu_\mu,\; h \nu_\mu \nu_\mu~, \label{eq:Hn4} \\  
H &\; \to \; e_4 \mu \; \to \; W\nu_\mu \mu, \; Z\mu\mu,\; h \mu\mu~, \label{eq:He4}
\end{align}
 which are also depicted in figure~\ref{fig:topologies}. In addition, $H$ could also decay into pairs of vectorlike leptons. This is however  limited to smaller ranges for masses in which these decays are kinematically open. Moreover, the final states are the same as in pair production of vectorlike leptons. We will not consider these possibilities here. Finally, although we  focus on the second family of SM leptons in final states, the modification for a different family of leptons or quarks is straightforward.

We show that in a large range of the parameter space branching ratios for the decay modes (\ref{eq:Hn4}) and (\ref{eq:He4}) can be sizable  or even dominant while satisfying constraints from searches for heavy Higgs bosons,  pair production of  vectorlike leptons~\cite{Dermisek:2014qca} obtained from searches for anomalous production of multilepton events and  constraints from precision electroweak (EW) observables~\cite{Agashe:2014kda}. Since the Higgs production cross section can be very large, for example the cross section for  a 200 GeV Higgs boson at 8 TeV (13 TeV)  LHC for $\tan \beta =1$ is 7pb (18pb)~\cite{Heinemeyer:2013tqa}, the final states above can be produced in large numbers. Thus searching for these processes could  lead to the simultaneous discovery of a new Higgs boson and a new lepton if they exist. Some of the decay modes in figure~\ref{fig:topologies} also allow for full reconstruction of the masses of both new particles in the decay chain.

The final states of the processes (\ref{eq:Hn4}) and (\ref{eq:He4}) are the same as final states of $pp \to WW, ZZ, Zh$ production or $H \to WW, ZZ$ decays with one of the gauge bosons decaying into second generation of leptons.  Since searching for leptons in final states is typically advantageous,  our processes contribute to a variety of existing searches.  Even searches for processes with fairly large cross sections can be significantly affected. For example, the contribution of  $pp \to H \to \nu_4 \nu_\mu \to W \mu \nu_\mu$  to $pp \to WW$ can be close to current limits while satisfying the constraints from  $H \to WW$. This has been recently studied in ref.~\cite{Dermisek:2015oja} in the two Higgs doublet model we consider here, and also in a more model independent way in ref.~\cite{Dermisek:2015vra}. However, the processes with tiny SM rates would be the best place to look for this scenario and here we will focus on such signatures. Examples of almost background free processes include $H \; \to \; h \nu_\mu \nu_\mu$ and $H \; \to \; h \mu\mu$ with $h\to \gamma \gamma$.

Vectorlike quarks and leptons near the electroweak scale provide a very rich phenomenology.  For example, similar  processes to (\ref{eq:Hn4}) and (\ref{eq:He4}) involving  SM-like Higgs boson decaying into $2\ell 2\nu$ or $4 \ell$ through a new lepton were previously studied in ref.~\cite{Dermisek:2014cia} and the  $4\ell$ case also in  ref.~\cite{Falkowski:2014ffa}. An explanation of the muon g-2 anomaly with vectorlike leptons was studied in \cite{Kannike:2011ng, Dermisek:2013gta}. Vectorlike quarks and possibly $Z'$  offer possibilities to explain anomalies in $Z$-pole observables~\cite{Choudhury:2001hs, Dermisek:2011xu, Dermisek:2012qx, Batell:2012ca}. Extensions of the SM with complete vectorlike families were considered to provide an understanding of  values of gauge couplings from IR fixed point behavior and threshold effects of vectorlike fermions~\cite{Dermisek:2012as, Dermisek:2012ke}.  Vectorlike quarks and leptons were also considered  in supersymmetric framework, see for example~refs.~\cite{Babu:1996zv, BasteroGil:1999dx, Kolda:1996ea, Barr:2012ma, Martin:2009bg, Bae:2012ir}. Further discussion and references can be found in a recent review~\cite{Ellis:2014dza}.

This paper is organized as follows. In section~\ref{sec:model} we briefly summarize the model, discuss constraints and present result for  branching ratios of the heavy Higgs boson  and new  leptons.  In section~\ref{sec:signatures} we discuss  relevant existing searches  and the most promising search strategies for each of the six processes. We summarize and present  concluding remarks in section~\ref{sec:conclusions}.

\section{Two Higgs doublet model with vectorlike leptons}
\label{sec:model}

In ref.~\cite{Dermisek:2015oja} we introduced an explicit model consisting of a type-II two Higgs doublet model augmented
by vectorlike pairs of new leptons: SU(2) doublets $L_{L,R}$,  SU(2) singlets $E_{L,R}$ and  SM singlets $N_{L,R}$, where the  $L_L$ and $E_R$ have the same  hypercharges as leptons in the SM. 
In order to avoid dangerous rates of lepton flavor changing transitions between the light leptons we  further assume that the new leptons mix only with one family of SM  leptons and we consider the mixing with the second family as an example. This can be achieved by requiring that the individual lepton number is an approximate symmetry (violated only by light neutrino masses). 
With these assumptions, one can write the most general renormalizable Lagrangian containing Yukawa and mass terms for the second generation of SM leptons and new vectorlike leptons.

After spontaneous symmetry breaking, $\left< H_u^0 \right> = v_u$ and $\left< H_d^0 \right> = v_d$ with $\sqrt{v_u^2 + v_d^2} = v = 174$ GeV (we also define $\tan \beta \equiv v_u / v_d$), the model can be summarized by mass matrices in the charged lepton sector, with left-handed fields $( \bar \mu_{L}, \bar L^-_L, \bar E_L )$ on the left and right-handed fields $( \mu_{R} , L^-_R, 
 E_R)^T$ on the right~\cite{Dermisek:2013gta},
\begin{eqnarray}
M_e \;  
&=& 
\begin{pmatrix}
 y_\mu v_d & 0 &  \lambda_E v_d\\
  \lambda_L v_d & M_L &  \lambda v_d\\
 0 & \bar \lambda v_d & M_E 
\end{pmatrix},
\end{eqnarray}
and in  the neutral lepton sector, with left-handed fields $( \bar \nu_\mu , \bar{L}_L^0 , \bar N_L)$ on the left and right-handed fields $( \nu_R = 0 ,
L_R^0 ,
N_R)^T$ on the right~\cite{Dermisek:2015oja},
\begin{eqnarray}
M_\nu  &=& 
\left( 
\begin{array}{ccc}
0 & 0 & \kappa_N v_u \\
0 & M_L & \kappa v_u \\
0 & \bar \kappa v_u & M_N \\
\end{array}
\right)~.
\label{eq:mm}
\end{eqnarray}
The superscripts on vectorlike fields represent the charged and the neutral components (we  inserted $\nu_R = 0$ for the  right-handed neutrino which is absent in our framework in order to keep  the mass matrix $3\times3$ in complete analogy with the charged sector).  The usual SM Yukawa coupling of the muon is denoted by $y_\mu$, the Yukawa couplings to $H_d$ are denoted by various $\lambda$s, the Yukawa couplings to $H_u$ are denoted by various $\kappa$s and finally the explicit mass terms for vectorlike  leptons are given by $M_{L,E,N}$. Note that explicit mass terms between SM and vectorlike fields (i.e. $\bar \mu_L L_R$ and $\bar E_L \mu_R$) can be rotated away. These mass matrices can be diagonalized by bi-unitary transformations and we label the two new charged and neutral mass eigenstates by $e_4, \, e_5$ and $\nu_4, \, \nu_5$ respectively:
\begin{eqnarray}
U_L^\dagger M_e U_R &=& {\rm diag}
\left(
 m_\mu,  m_{e_4} ,  m_{e_5}
\right)~, \\
V_L^\dagger M_\nu V_R &= &  {\rm diag} 
\left(0, m_{\nu_4} , m_{\nu_5}
\right)~.
\label{eq:unitary}
\end{eqnarray}

Since SU(2) singlets mix with SU(2) doublets, the couplings of all involved particles to the $Z$, $W$ and Higgs bosons are in general modified. The flavor conserving couplings receive corrections and flavor changing couplings between the muon (or muon neutrino)  and heavy leptons are generated. The relevant formulas for these couplings in terms of diagonalization matrices defined above can be found in Refs.~{\cite{Dermisek:2015oja, Dermisek:2013gta}. In the limit of small mixing, approximate analytic expressions  for diagonalization matrices can be obtained which are often useful for understanding of numerical results. These are also given in Refs.~{\cite{Dermisek:2015oja, Dermisek:2013gta}.

\subsection{Branching ratios of the heavy Higgs boson}
\label{sec:brH}

In order to focus on decay modes (\ref{eq:Hn4}) or (\ref{eq:He4}) we either allow mixing only in the neutral sector, $\kappa$ couplings, or only in the charged sector, $\lambda$ couplings.  
We further assume that the relevant lighter new lepton, $\nu_4$ or $e_4$, is heavier that $m_H / 2 $ to avoid decays into pairs of new leptons and the heavier new lepton, $\nu_5$ or $e_5$, is heavier than $H$.  Finally, we work in the limit with the light Higgs boson being fully SM-like and thus the heavy CP even Higgs $H$ has no direct couplings to gauge bosons.
We apply these requirements on randomly generated points in the parameter space specified by following ranges:
\begin{align}
m_H &\in [130, 340]~{\rm GeV}~,\\
\tan\beta &\in [0.3, 3]~, \\
\kappa_N, \kappa, \bar \kappa ~~{\rm or}~~ \lambda_L, \lambda_E, \lambda, \bar \lambda &\in [-0.5, 0.5]~, \\
M_{L,N,E} &\in [100, 500]~{\rm GeV}~,
\end{align}
where we focus on $m_H < 2 m_t$ in order to avoid $H\to t\bar t$ and on the small $\tan\beta$ region where the heavy Higgs production cross section is the largest.

We impose constraints from precision EW data related to the muon and muon neutrino: muon lifetime, $Z$-pole observables ($Z$ partial width to $\mu^+\mu^-$, the invisible width, forward-backward asymmetry, left-right asymmetry) and the $W$ partial width; constraints from oblique corrections, namely from S and T parameters; and the LEP limit on the mass of a new charged lepton, 105 GeV. These constraints are obtained from~ref.~\cite{Agashe:2014kda}. We also impose constraints on pair production of  vectorlike leptons~\cite{Dermisek:2014qca} obtained from searches for anomalous production of multilepton events.\footnote{For the discussion of prospects of multilepton searches for vectorlike leptons in the case of mixing with the 3rd generation of SM leptons, see ref.~\cite{Kumar:2015tna}.}

In addition to constraints on new leptons we also impose constraints from searches for new Higgses: $H \to WW$ \cite{CMS:bxa,Chatrchyan:2013iaa} and $H \to \gamma \gamma$ \cite{CMS:2014onr}. Although the processes in (\ref{eq:Hn4}) or (\ref{eq:He4}) do not contribute to $H \to WW$ directly, they contribute to the same final states as obtained from decays of $W$ bosons. 
 In applying the constraints from $H \to WW$ we follow the analyses presented in Refs.~\cite{Dermisek:2013cxa,Dermisek:2015vra}. We implement the cut-based analysis using the data for $e\mu \nu_e \nu_\mu$ final state in \cite{CMS:bxa,Chatrchyan:2013iaa} but the results using $\mu \mu \nu_\mu \nu_\mu$ would be similar. 
 Since our $H$ has no direct coupling to the $W$ boson, it is only the top quark and new charged leptons in loops that contribute to  $H \to \gamma \gamma$. The top quark contribution to $H \to \gamma \gamma$ scales as $\cot^2 \beta$ and thus this process is highly constraining the parameter space at very small $\tan \beta$. In the usual two Higgs doublet model this was studied in ref.~\cite{Song:2014lua}. However, new charged leptons can  enhance or partially cancel the contribution from the top quark depending on the signs of the new Yukawa couplings. Thus the allowed range of $\tan\beta$ is expected to extend to lower values compared to the usual two Higgs doublet model. Since new leptons also couple to the SM-like Higgs boson, the constraints from $h \to \gamma \gamma$ have to be also satisfied. These at present still allow for a sizable new physics contribution~\cite{Khachatryan:2014ira}. 

\begin{figure}
\begin{center}
\includegraphics[width=0.49\linewidth]{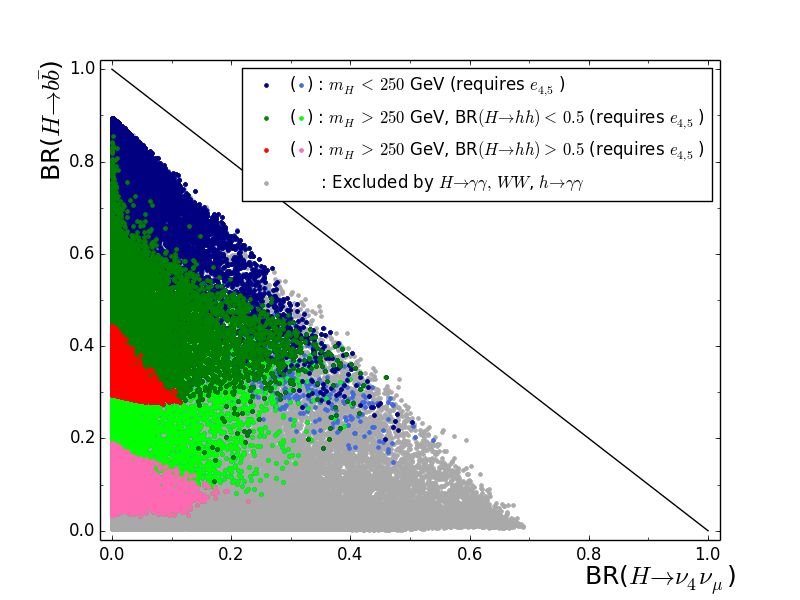}
\includegraphics[width=0.49\linewidth]{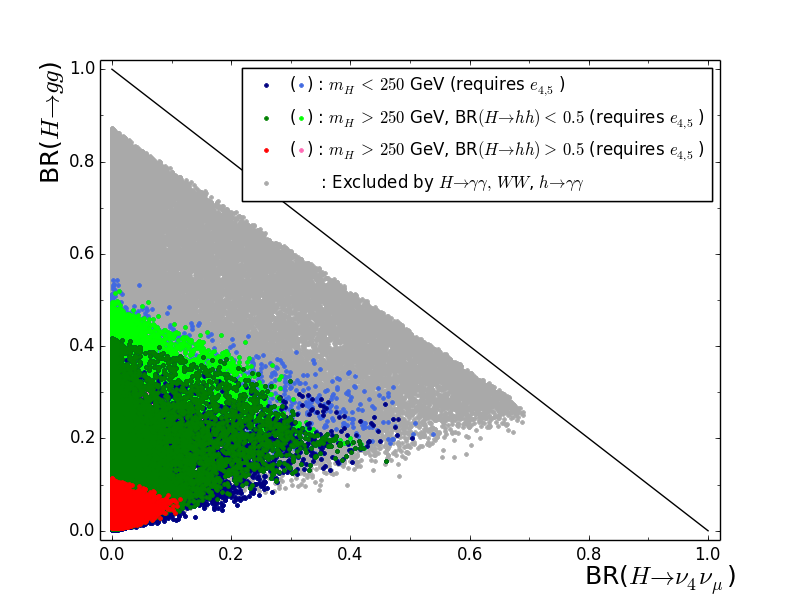}
\includegraphics[width=0.49\linewidth]{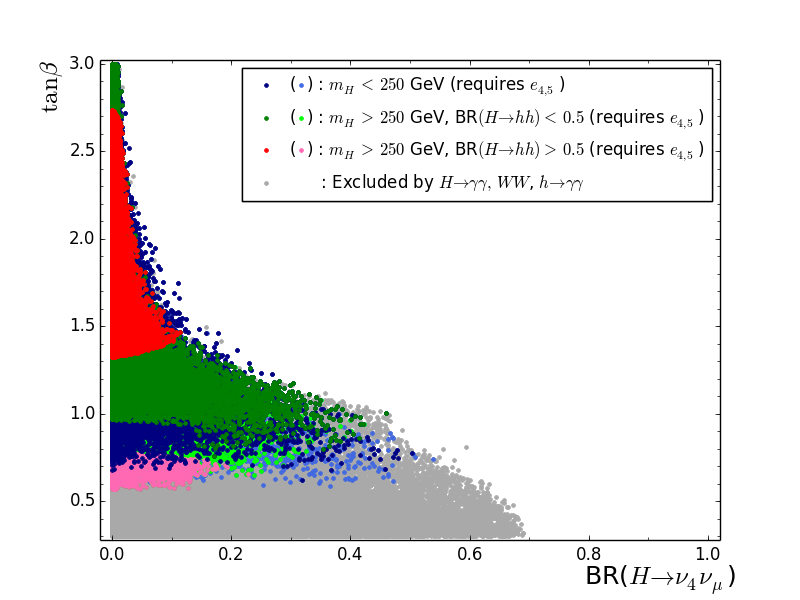}
\includegraphics[width=0.49\linewidth]{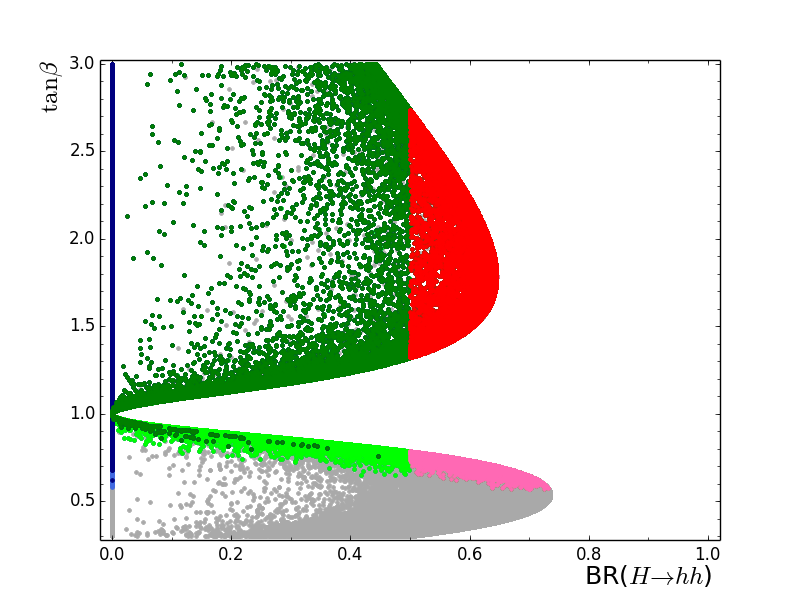}
\caption{Heavy Higgs branching ratios in the case of mixing in the neutral sector allowing the $H \to \nu_4 \nu_\mu$ decay. All the points satisfy experimental constraints from precision EW data and the limits on pair production of vectorlike leptons at the LHC. The dark colored points (non-gray points) additionally satisfy the bounds from $H \to \gamma \gamma,\, WW$ and $h \to \gamma \gamma$. Note that contributions to $h\to\gamma\gamma$ and $H\to \gamma\gamma$ are possible only if mixing in the charged lepton sector is simultaneously allowed. Various colors indicate different ranges of $m_H$ and  BR($H \to hh$) specified in the legend. }
\label{fig:HBR}
\end{center}
\end{figure}
\begin{figure}
\begin{center}
\includegraphics[width=.49\linewidth]{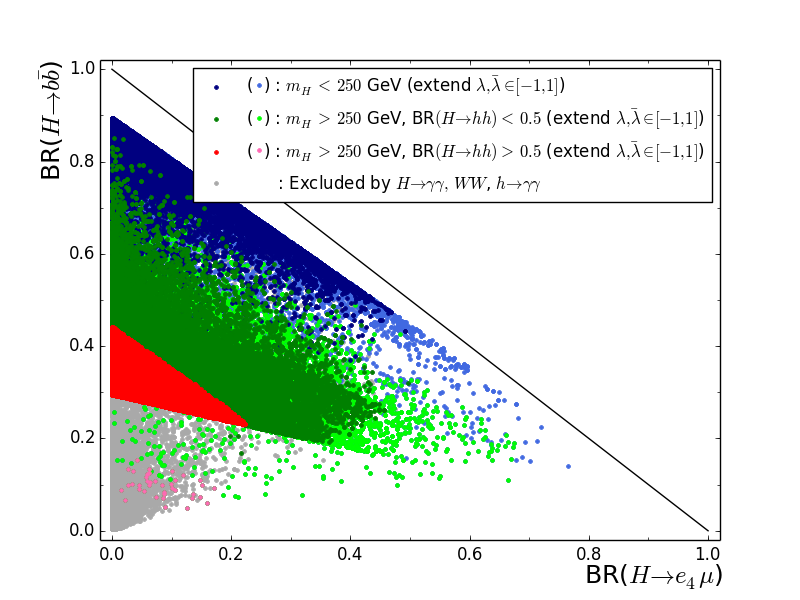}
\includegraphics[width=.49\linewidth]{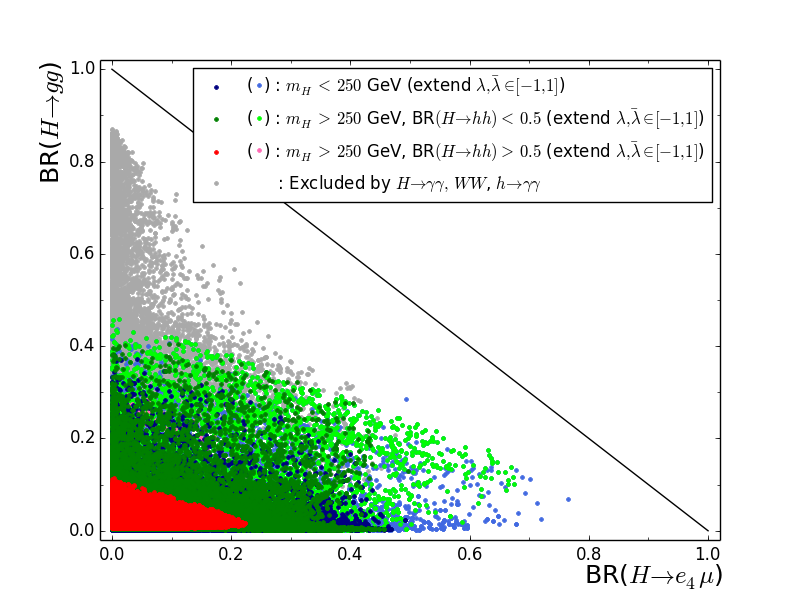}
\includegraphics[width=.49\linewidth]{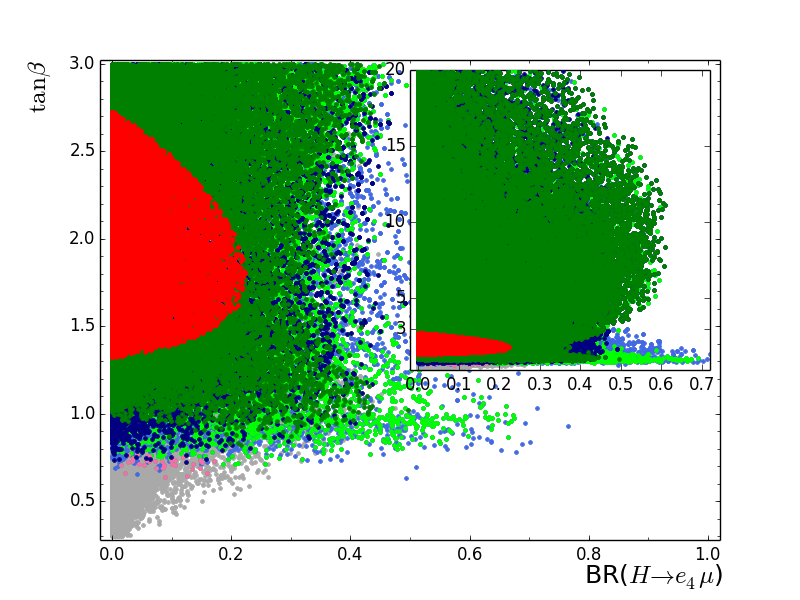}
\includegraphics[width=.49\linewidth]{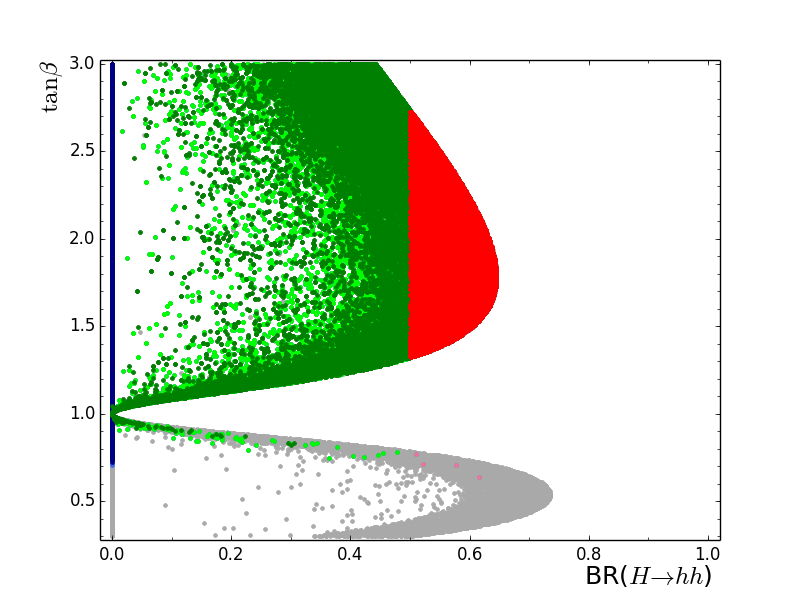}
\caption{Branching ratios of $H$ in the case of mixing in the charged sector allowing the $H \to e_4 \mu$ decay. Color coding is the same as in figure~\ref{fig:HBR}. We inserted the subfigure extending $\tan\beta$ up to 20 in the lower-left plot.
}
\label{fig:HBRe}
\end{center}
\end{figure}

Results of the scan in the case of mixing in the neutral lepton sector, allowing for $H\to \nu_4 \nu_\mu$, are depicted in figure~\ref{fig:HBR} in various planes of relevant branching ratios and $\tan\beta$. The dark colored points satisfy all the limits summarized above. The gray points satisfy all the limits on new leptons but are excluded by $H \to \gamma \gamma,\, WW$ and $h \to \gamma \gamma$. Finally, the light colors represent the points which are phenomenologically viable and satisfy $H \to \gamma \gamma$ and $h \to \gamma \gamma$ limits only if mixing in the charged lepton sector is simultaneously allowed which partially cancels the contribution from the top quark. In this case, for simplicity, we  allow  $L$ and $E$ to mix but not with the muon and we conservatively extend the ranges for $\lambda $ and $ \bar \lambda$ to [-1,1]. We clearly see the lower bound on $\tan\beta \simeq 0.55$. Without the mixing in the charged sector  the lower limit on $\tan \beta $ moves to about 0.7.

Similarly, results of the scan in the case of mixing only in the charged  lepton sector, allowing for $H\to e_4 \mu$, are given in figure~\ref{fig:HBRe} in the same planes and color scheme. 
In this case, the $\tan\beta$ dependence is not so significant because the new Yukawa coupling inducing $H \to e_4 \mu$ scales with $\tan\beta$ in the same way as the $b \bar b$ coupling. The lowest possible value of $\tan\beta$ is 0.7 as it was in the $H \to \nu_4 \nu_\mu$ case. Even extending 
the ranges for $\lambda $ and $ \bar \lambda$ to [-1,1], as indicated by light colors, only allows $\tan \beta \gtrsim 0.6$.

For completeness we plot the same points in the plane of Higgs branching ratios for $H\to \nu_4 \nu_\mu$ and $H\to e_4 \mu$ versus the Higgs mass in figure~\ref{fig:BRHmH}. Finally, although we focussed on the CP even Higgs, the results for CP odd Higgs would be qualitatively similar.

\subsection{Branching ratios of the lightest neutral and charged leptons}
\label{sec:br4}

The branching ratios of new lightest neutral and charged leptons decaying through $ Z$ and $W$ bosons are shown in figure~\ref{fig:vllBR}. Although not explicitly shown, the remaining branching ratio of the decay through the SM Higgs boson can be easily read out since only three decay modes are possible. In the left panels only constraints from EW precision data and direct searches are imposed. In the right panels we include the impact of constraints from searches for anomalous production of multilepton events~\cite{Dermisek:2014qca}. In these plots, the colors indicate the doublet fractions of $\nu_4$ or $e_4$. The blue, cyan, magenta, and red points have doublet fractions  in the ranges  [95,100]\%, [50,95]\%, [5, 50]\%, and  [0, 5]\% respectively. The doublet fraction the $\nu_a \; (a=4,5)$ is defined as
\begin{align}
\frac{1}{2} \left\{ \left[ (V_L^\dagger)_{a2}\right]^2 +\left[ (V_L^\dagger)_{a4}\right]^2 +\left[ (V_R^\dagger)_{a4}\right]^2 \right\}~
\end{align}
and the doublet fraction of $e_a \; (a=4,5)$ is obtained by replacing the $V_{L,R}$ matrices by $U_{L,R}$ matrices in the formula above.
Singlet fractions are given by $1$ -- (doublet  fraction). 

We see that the multilepton searches are very constraining for doublet-like new leptons and do not even allow a doublet-like  charged lepton in the mass range considered. Note however that in the case of charged leptons with large $BR(e_4\to W\nu) $ the constraints come from the pair production of  $\nu_4$ that  accompanies doublet-like $e_4$ or from $e_4\nu_4$ production; the $e_4 e_4$ pair production is not directly constrained by multilepton searches. Without mixing in the neutral sector, $BR(\nu_4\to W\mu) =1 $ and this decay mode is highly constrained~\cite{Dermisek:2014qca}. Allowing simultaneously full mixing in both charged and neutral sectors would relax this constraint somewhat.

The main features of plots in figure~\ref{fig:vllBR} can be understood from analytic formulas for couplings that can be obtained in the limit of small mixing~\cite{Dermisek:2015oja}. For singlet-like $e_4$ or $\nu_4$, the flavor changing couplings to W and Z have, in the leading order,  the same dependence on parameters controlling the mixing. Thus this dependence  disappears in the ratio and we find  $\Gamma(e_4 \to W \nu_\mu)/\Gamma(e_4 \to Z \mu)$  to be approximately 2:1 leading to the red bands with this slope.

For doublet-like $\nu_4$, the  flavor changing couplings to $A = W,Z,h$ have the form $\kappa_N (\alpha_A \kappa + \beta_A \bar\kappa)$ where $\alpha_A$ and $ \beta_A$ are functions of $\tan\beta$, $M_N$ and $M_L$. This implies immediately that, for fixed $\tan\beta$, $M_N$ and $M_L$, a scan over the three couplings $\kappa_N$, $\kappa$ and $\bar\kappa$ will result in an ellipse in the ${\rm BR} (\nu_4 \to Z\nu_\mu) - {\rm BR} (\nu_4 \to W\mu)$ plane. The major axis of these ellipses is almost horizontal for $M_L \gtrsim v$; for smaller $M_L$ the ellipses collapse onto the diagonal, which corresponds to ${\rm BR} (\nu_4 \to h\nu_\mu) \sim 0$. This behavior can be seen in figures 3, 4 and 7 of ref.~\cite{Dermisek:2015oja} for various choices of $m_{\nu_4}$. 

Finally, for the doublet-like $e_4$, there are two couplings, $\lambda_L$ and $\lambda_E$,  connecting the muon to vectorlike fermions and thus controlling the overall strengths of flavor changing couplings to $W$, $Z$, and $h$. In result, there is significantly more freedom and generating couplings to $W$, $Z$, and $h$  are less correlated.  

\begin{figure}
\begin{center}
\includegraphics[width=.49\linewidth]{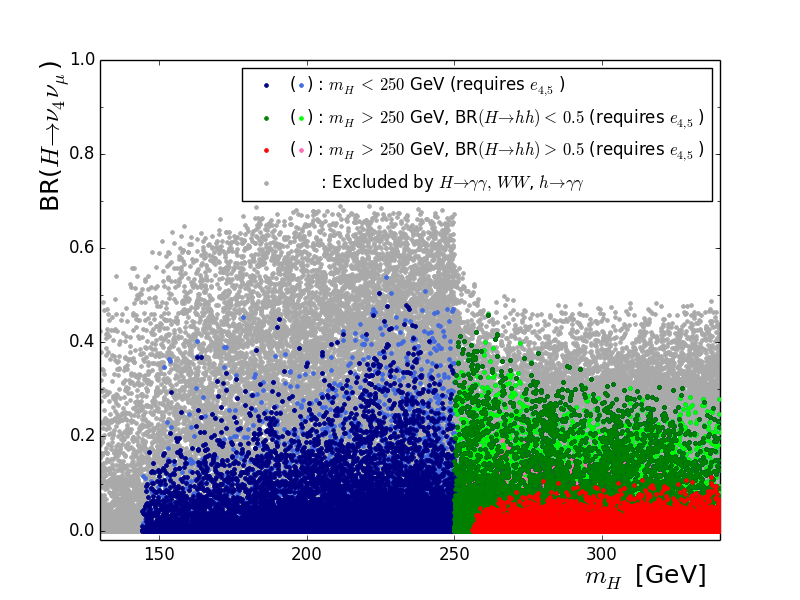}
\includegraphics[width=.49\linewidth]{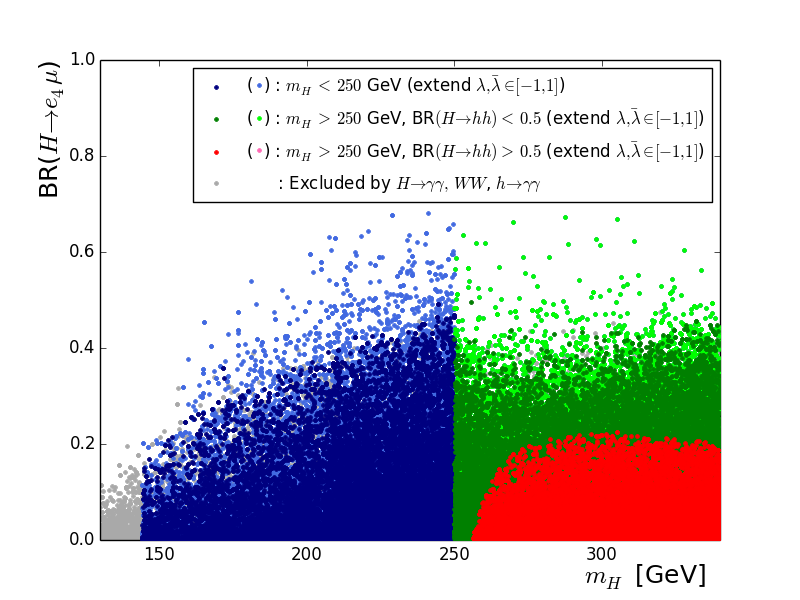}
\caption{
Points from figure~\ref{fig:HBR} plotted in  $m_H$ -- BR($H \to \nu_4 \nu_\mu$) and $m_H$ -- BR($H \to e_4 \mu$) planes. 
}
\label{fig:BRHmH}
\end{center}
\end{figure}
\begin{figure}
\begin{center}
\includegraphics[width=.49\linewidth]{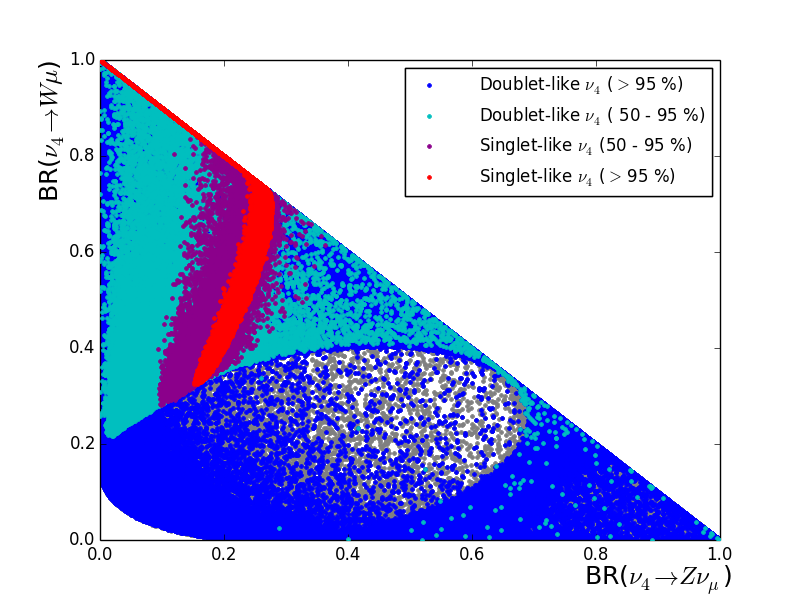}
\includegraphics[width=.49\linewidth]{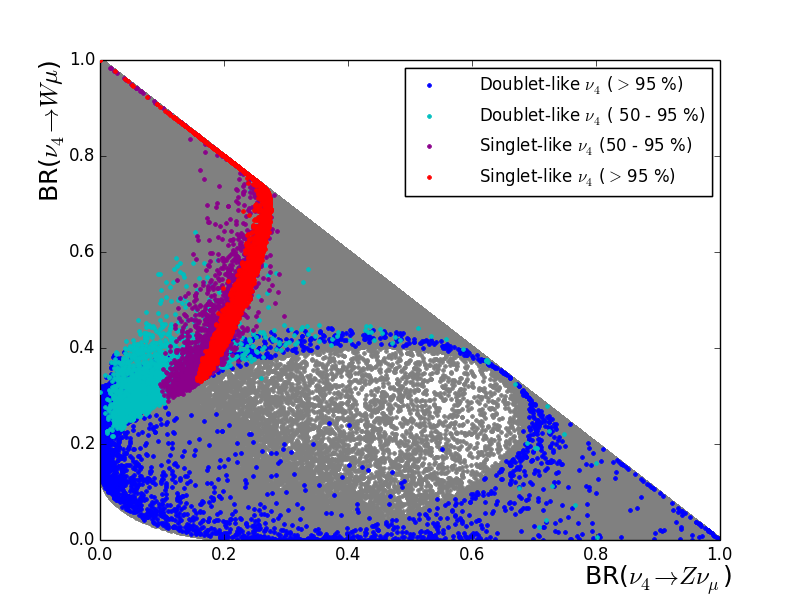}
\includegraphics[width=.49\linewidth]{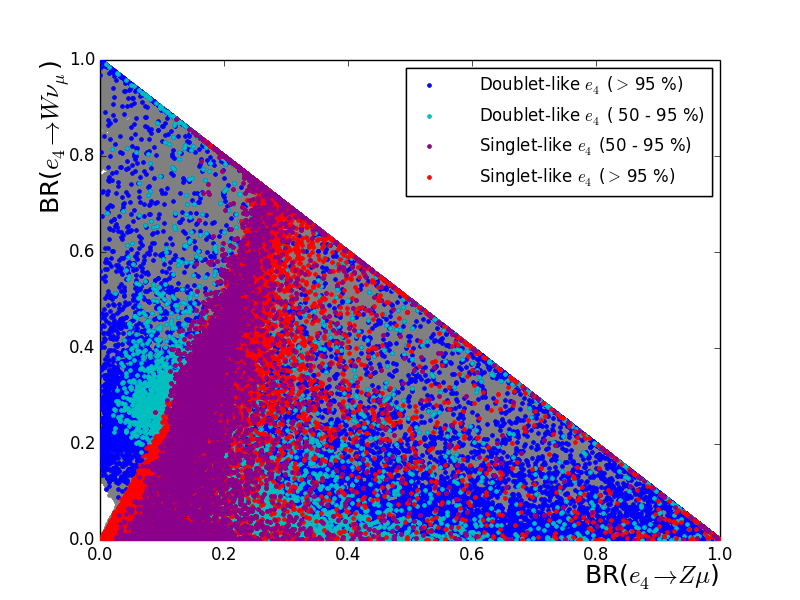}
\includegraphics[width=.49\linewidth]{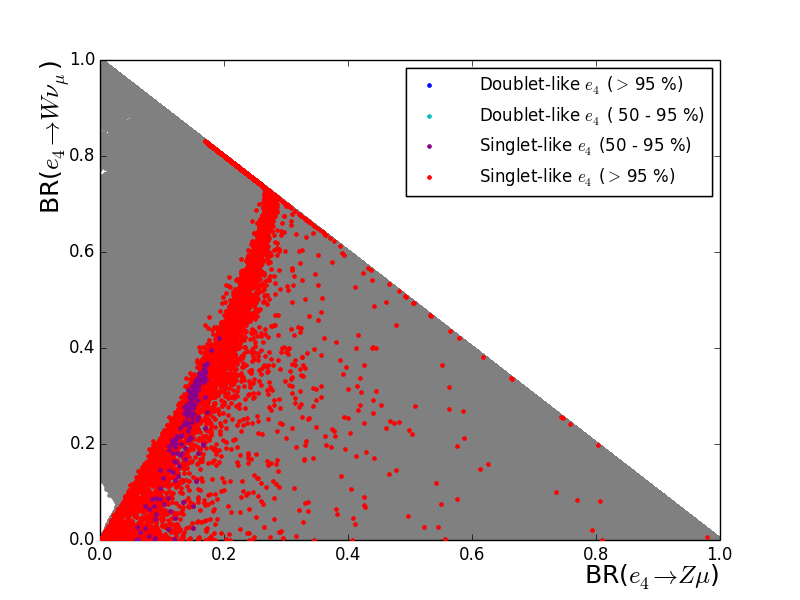}
\caption{
Branching ratios of $\nu_4$ and $e_4$. Gray points in the left plots are ruled out by constraints from precision EW data and in the right plots also by searches for anomalous production of multilepton events. The blue, cyan, magenta, and red points have doublet fractions in the ranges [95,100]\%, [50,95]\%, [5, 50]\%, and  [0, 5]\% respectively.
}
\label{fig:vllBR}
\end{center}
\end{figure}

In order to illuminate  more subtle   features of the scenario, we plot the same points in the planes of branching ratios versus the masses of $\nu_4$ and $e_4$ in figure~\ref{fig:BRmHvll}.

\begin{figure}
\begin{center}
\includegraphics[width=.49\linewidth]{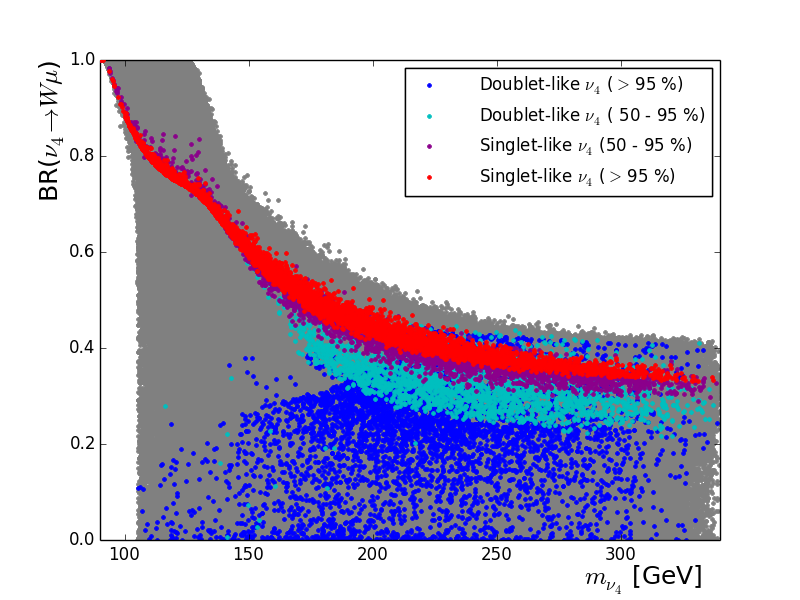}
\includegraphics[width=.49\linewidth]{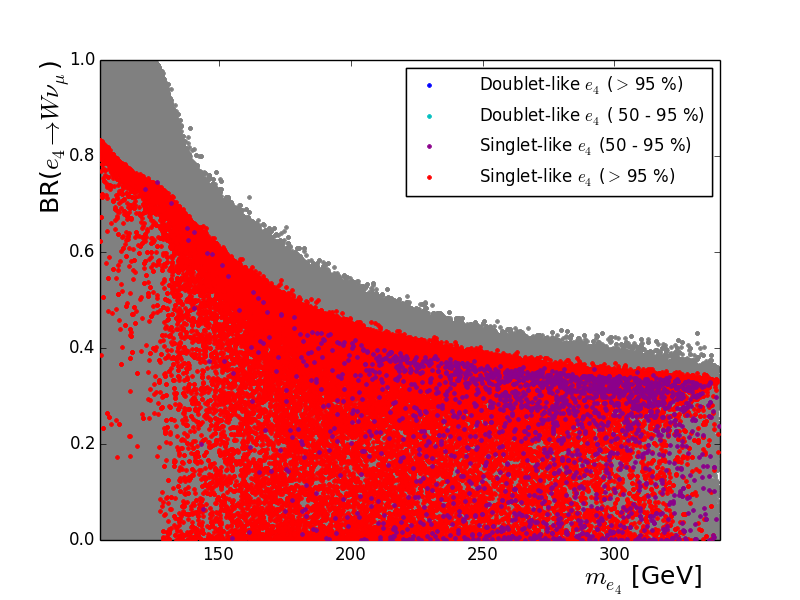}
\includegraphics[width=.49\linewidth]{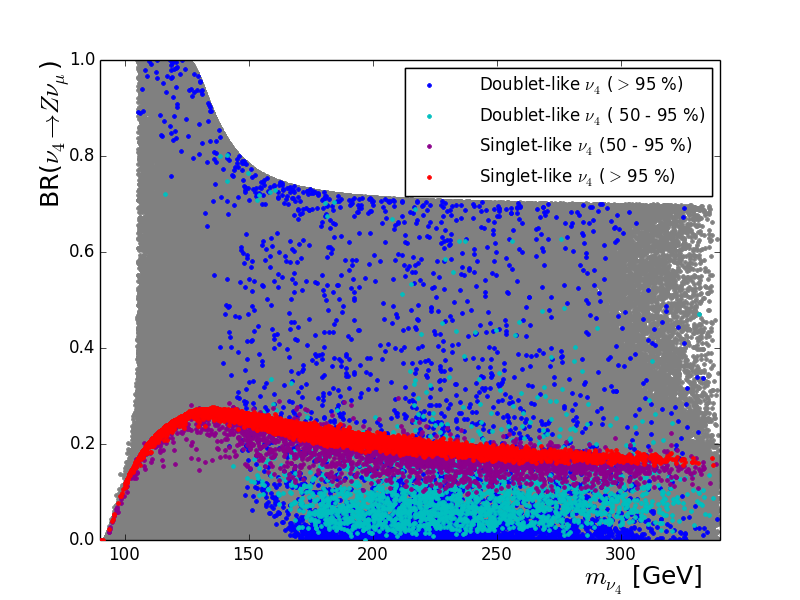}
\includegraphics[width=.49\linewidth]{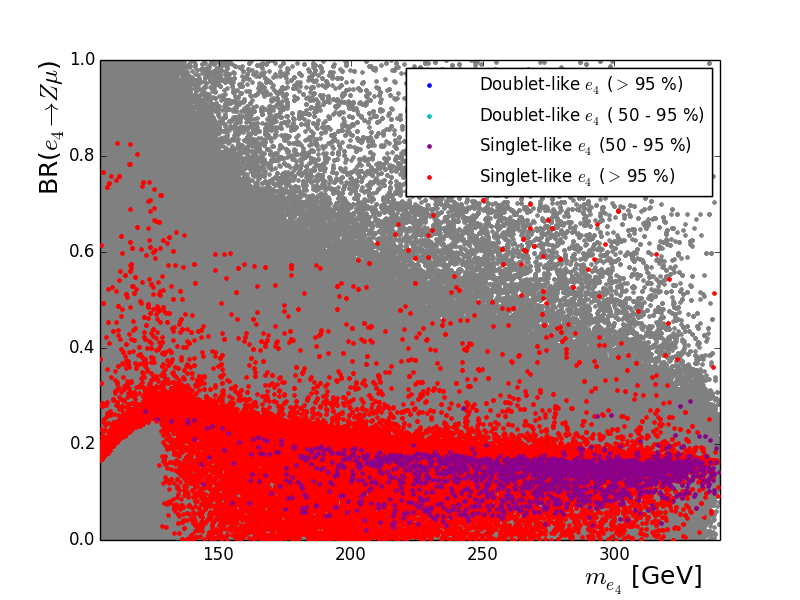}
\includegraphics[width=.49\linewidth]{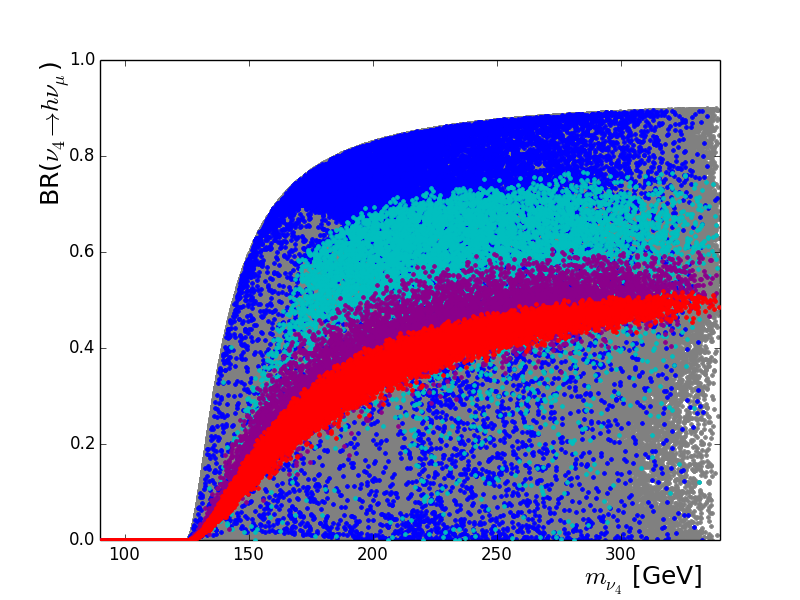}
\includegraphics[width=.49\linewidth]{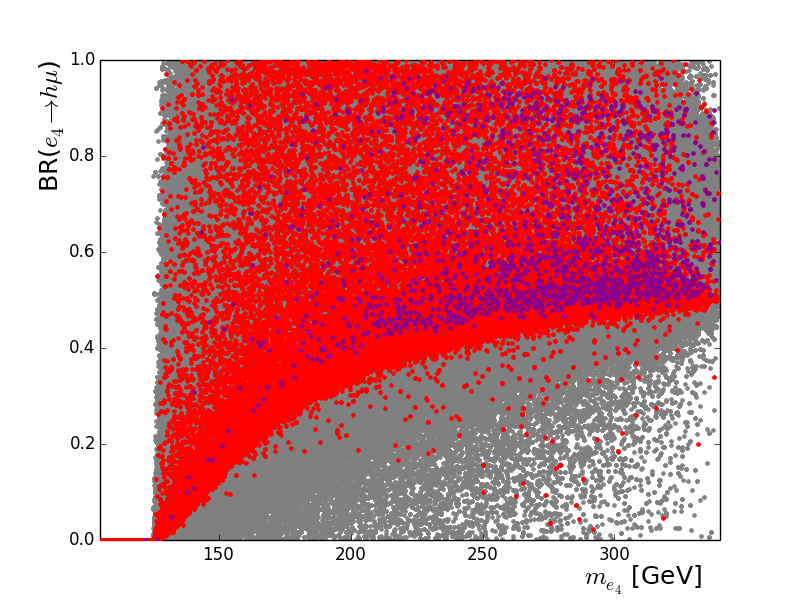}
\caption{
Points from figure~\ref{fig:vllBR} plotted in planes of branching ratios and corresponding lightest new lepton mass.}
\label{fig:BRmHvll}
\end{center}
\end{figure}
%

\section{Signatures}
\label{sec:signatures}
In this section we discuss each of the novel heavy Higgs decay modes with details of the main features of each channel, of existing experimental searches to which these new process contribute, and of possible new searches. Considering the variety of possible final states, detailed Monte Carlo studies of these signatures are beyond the scope of this paper.

For estimates of rates of various processes we will use, as a reference point,  the production cross section of a 200 GeV Higgs boson at 8 TeV  LHC for $\tan \beta =1$ which is 7pb. The corresponding  cross section at 13 TeV  LHC is 18pb, and for different values of $\tan \beta$ these numbers should be divided by $\tan^2 \beta$. Furthermore, we will assume ${\rm BR} (H\to \nu_4\nu)$ or ${\rm BR} (H\to e_4\mu)$ to be 50\% and branching ratios of $\nu_4$ or $e_4$ relevant for a given process to be 100\%. These branching ratios are close to the upper possible values allowed as we saw in the previous section. Note however that after imposing all constraints $e_4\to Z\mu$ branching ratios above 40\%, while possible, are difficult to achieve.

\subsection{$H\to W \mu\nu_\mu$}
\label{sec:Wmunu}
The diagrams in figures~\ref{fig:topologies}a and \ref{fig:topologies}d yield the $W\mu\nu_\mu$ final state. A detailed study of these topologies with $W\to \ell\nu \; (\ell=e,\mu)$ has been presented in Refs.~\cite{Dermisek:2015vra, Dermisek:2015oja}, where with focus on contributions to the existing $pp\to WW$ and $pp\to H\to WW$ measurements. 

A crucial feature of the process in figure~\ref{fig:topologies}a is that the intermediate $\nu_4 \to W\mu$ decay, with a hadronically decaying $W$, allows the kinematic reconstruction of the neutral fermion $\nu_4$. Experimental studies of this mode can be affected by searches for semileptonic decays of a heavy Higgs ($H\to WW\to \ell\nu jj$) which have been presented in Refs.~\cite{ATLAS:2011ae, Chatrchyan:2013yoa, CMS:2015lda, ATLAS_WW8, ATLAS_WW13}. In these searches, the assumption that the observed missing transverse momentum is caused by a neutrino emitted in the $W$ decay, is used to reconstruct the complete four-momentum of the neutrino; thus allowing the reconstruction of the Higgs mass ($m_H^2 = (p_\ell + p_\nu + p_{jj})^2$). The efficiency loss due to this reconstruction procedure is about 50\%. However, in our case, the neutrino does not originate from a $W$ and this procedure does not reconstruct the correct Higgs mass.  An alternative approach is to consider the transverse mass $m_T = (\slashed{E}_T^2 - \slashed{p}_T^2)^{1/2}$ that is expected to have an edge at $m_H$. Moreover the $p_T$ distribution of the neutrino is expected to be different because the latter does not originate from a $W$. Finally, for Higgs masses above 200 GeV our signal is potentially enhanced, with respect to the one studied in ref.~\cite{ATLAS:2011ae}, by the ratio of ${\rm BR} (H\to \nu_4\nu \to W\mu\nu \to jj\mu\nu) \lesssim 0.5 \times 1 \times 0.7 \sim 0.35$ (where the maximal $H$ and $\nu_4$ branching ratios are taken from figure~\ref{fig:BRHmH} and \ref{fig:BRmHvll}) to ${\rm BR} (H\to WW\to jj\mu\nu) = 2 \times 0.74 \times 0.7 \times 0.1 \simeq 0.1$ (where we included a combinatoric factor of 2 for the $W$'s decays). For $W\to \tau\nu$ our process is $H \to \tau \mu \nu\nu$ and is constrained by searches for $H\to \tau \tau$ (with a leptonic tau) and $H \to \tau \mu$.

\subsection{$H\to Z \nu_\mu\bar\nu_\mu$}
\label{sec:Znunu}
The diagram in figure~\ref{fig:topologies}b leads to the $Z\nu\bar\nu$ final state. For $Z\to \nu\bar\nu$ our process contributes to the jet plus missing $E_T$ signature, that is common to monojet searches for dark matter pair production or for invisible Higgs boson decays. For instance, in Refs.~\cite{Aad:2015zva, Khachatryan:2014rra} the ATLAS and CMS collaborations placed upper limits on the visible cross section defined as the product of cross section, Monte Carlo acceptances and  detector efficiencies, i.e. the observed number of events divided by the integrated luminosity. Requiring $\slashed{E}_T > 150 \; {\rm GeV}$, the ATLAS limit is $726 \; {\rm fb}$ which is larger than the typical values that we can obtain. For our reference point, even before requiring an extra high-$p_T$ jet and including acceptances and detector efficiencies, the total cross section is $\sigma(pp\to H) \times {\rm BR} (H\to \nu_4\nu_\mu \to Z \nu_\mu \nu_\mu \to \nu_\mu \nu_\mu \nu_\ell \nu_\ell) \lesssim (7 \;{\rm pb}) \times 0.5 \times 1 \times 0.2 \sim 0.7 \; {\rm pb}$. 

For leptonic $Z$ decays, $Z\to \ell\ell$, our signal contributes to the $pp\to ZZ\to \ell \ell \nu \bar\nu$ measurement~\cite{Khachatryan:2015pba} and to searches for $H\to ZZ \to \ell \ell \nu \bar\nu$~\cite{Aad:2015kna}, mono-$Z$ with missing energy, $Z + \slashed{E}_T$~\cite{Aad:2014vka, CMS:2015kha} and $Z h (H) \to \ell\ell + \slashed{E}_T $~\cite{Aad:2014iia}. For instance, the CMS search for $Z + \slashed{E}_T$~\cite{CMS:2015kha} finds a limit on the visible cross section (for various $\slashed{E}_T$ cuts) at the 1-2 fb level; in our case the $m_H = 200$ GeV total cross section is up to $\sigma(pp\to H) \times {\rm BR} (H\to \nu_4\nu_\mu \to Z\nu_\mu \nu_\mu \to \ell\ell\nu_\mu \nu_\mu) \lesssim (7 \;{\rm pb}) \times 0.5 \times 1 \times 0.06 \sim 210 \; {\rm fb}$, before acceptances and efficiencies. Therefore, this channel is likely to offer the strongest constraint. 

The hadronic $Z\to jj$ mode is less clean and has been studied in the context of the $H\to ZZ \to jj \nu \bar\nu$ presented in ref.~\cite{Aad:2015kna}.

\subsection{$H\to Z\mu\mu$}
\label{sec:Zmumu}
The diagram in figure~\ref{fig:topologies}e leads to a $Z\mu\mu$ final state with two resonances corresponding to the Higgs, $P_H = p_Z + p_{\mu_1} + p_{\mu_2}$, and charged vectorlike lepton, $P_{e_4} = p_Z + p_{\mu_1}$ (up to a dilution due to picking the wrong muon).  

The most promising channel is $H\to Z\mu\mu\to \ell\ell\mu\mu$. A recent ATLAS search for $e_4 \to Z \ell \to 3 \ell$ with the $e_4$ produced with another vectorlike lepton in a Drell-Yan process is presented in ref.~\cite{Aad:2015dha} where the visible cross section into $4\ell$, $3\ell+jj$ and $3\ell$ final states is found to be smaller than 1 fb. For our reference point, we obtain $\sigma(pp\to H) \times {\rm BR} (H\to e_4\mu\to Z\mu\mu \to \ell\ell \mu\mu)  \lesssim (7 \;{\rm pb}) \times 0.5 \times 1 \times 0.06 \sim 210 \; {\rm fb}$. Since our signal has a different underlying process and kinematics with respect to vectorlike leptons pair production, we expect our acceptance for the ATLAS search to be somewhat smaller than the quoted $(20-50)\%$ acceptance. In addition, one could additionally search for a fourth lepton and require one lepton pair to form a $Z$ while imposing a $Z$ veto on the second pair thus suppressing $ZZ$ backgrounds; the four leptons invariant mass is also expected to peak at the heavy Higgs mass. 

In ref.~\cite{Khachatryan:2015cwa}, CMS presented a study of a heavy SM-like Higgs decaying to $ZZ$ and found that for Higgs masses smaller than about 500 GeV the 95\% upper limit on the total cross section into four leptons is $0.1 \times \sigma_{\rm SM} (pp\to H \to ZZ \to 4\ell)$. For $m_H=200\; {\rm GeV}$ this corresponds to about $0.1 \times (7 \; {\rm pb}) \times 0.255 \times (3 \times 0.03^2) \sim 0.48 \; {\rm fb}$ (the factor of 3 takes into account that only the $4\mu$ and $2e2\mu$ final states should be included), which has to be compared to our reference cross section of about 210 fb discussed above. Since these searches require two on-shell $Z$ bosons compared to the single $Z$ in our signal, we expect the Monte Carlo level acceptances for this process to be very small. A similar argument applies to $pp\to ZZ\to 4\ell$~\cite{CMS:2014xja}. However, one can again impose a $Z$ veto on two of the charged leptons, almost completely removing the SM background.

The $Z\to jj$ mode is also interesting but it yields experimental bounds that are roughly an order of magnitude weaker than in the di-lepton case~\cite{Khachatryan:2015cwa}. The invisible $Z$ channel is problematic because it leads to a dimuon plus missing energy final state in which the two muons do not reconstruct a $Z$. This final state would also contribute to a $WW$ search. 

\subsection{$H\to h\mu\mu$}
\label{sec:hmumu}
This mode stems from the diagram in figure~\ref{fig:topologies}f and leads to large contributions to several very promising final states: $(\gamma\gamma)_h \mu\mu$, $(ZZ^*)_h \mu\mu \to 6\ell$, $(WW^*)_h \mu\mu \to 4\ell + \slashed{E}_T$ and $(b\bar b)_h \mu\mu$ (the subscript indicates the particles whose invariant mass reconstruct the SM Higgs). Standard Model backgrounds to the first three modes are essentially absent making them golden channels to discover this process. In fact, the dominant SM backgrounds are $pp \to h (\gamma^*,Z)$; moreover, the latter can be further suppressed by requiring the $Z$ to be virtual by vetoing di-leptons with invariant mass close to $m_Z$. 

For our reference point we find $\sigma(pp\to H) \times {\rm BR} (H\to e_4\mu\to h \mu \mu\to \gamma\gamma\mu\mu) \lesssim (7 \;{\rm pb}) \times 0.5 \times 1 \times 3 \cdot 10^{-3} \sim 10.5 \; {\rm fb}$ and $\sigma(pp\to H) \times {\rm BR} (H\to e_4\mu\to h \mu\mu\to Z Z^* \mu\mu\to 4\ell 2\mu) \lesssim (7  \;{\rm pb}) \times 0.5 \times 1 \times 1.2\cdot 10^{-4} \sim 0.4 \; {\rm fb}$. A large number of expected events (over almost zero background) could be already present in the existing data set.

\subsection{$H\to h\nu\bar\nu$}
\label{sec:hnunu}
The SM Higgs plus missing energy signal, depicted in figure~\ref{fig:topologies}c, is also very interesting because it overlaps with dark matter searches. 

The clear golden mode is the $(\gamma\gamma)_h + \slashed{E}_T$ final state. A search for this signal has been performed in ref.~\cite{Aad:2015yga} where a small excess of 3 events has been observed over (essentially) no background. In fact, the $pp\to Zh\to \nu\nu\gamma\gamma$ process has a total cross section of about 0.2 fb and, with a 10\% acceptance, leads to about 0.5 events at 20 fb${}^{-1}$. The cross section for our reference point is about $\sigma(pp\to H) \times {\rm BR} (H\to \nu_4 \nu_\mu\to h \nu_\mu\nu_\mu \to  \gamma\gamma\nu_\mu \nu_\mu) \lesssim (7  \;{\rm pb}) \times 0.5 \times 1 \times 3\cdot 10^{-3} \sim 10.5  \; {\rm fb}$. Assuming that our signal acceptance is identical to the acceptance used in this search (10\%), we expect about 21 events with 20 fb${}^{-1}$ of integrated luminosity. Future updates of this search will certainly place interesting constraints on our model.

In ref.~\cite{Aad:2015uga} the $pp\to (Z\to b\bar b) (h\to {\rm invisible}) \to b\bar b + \slashed{E}_T$ was studied. Interestingly a small excess of 20 events is observed for $m_{bb} \sim 125 \; {\rm GeV}$ that would be compatible with our signal with $h\to b\bar b$. For our reference point and considering the $h\to b\bar b$ decay of the SM Higgs we get a cross section of up to $\sigma(pp\to H) \times {\rm BR} (H\to \nu_4 \nu_\mu\to h \nu_\mu \nu_\mu\to b b \nu_\mu\nu_\mu) \lesssim (7 \;{\rm pb}) \times 0.5 \times 1 \times 0.57\sim 2.0 \; {\rm pb}$. This is further reduced by the $b$-tagging efficiency ($0.7^2 = 0.5$)  and by the acceptance. In ref.~\cite{Chatrchyan:2014tja}, CMS presents a similar study but the di-jet invariant mass distribution after $b$-tagging is not presented; hence we do not know whether an excess at $m_{bb} = 125 \; {\rm GeV}$ is seen. 

In Refs.~\cite{Aad:2014xzb} ATLAS performed a dedicated search for $pp\to (h\to b\bar b) Z/W$. Events are classified according to the number of leptons in the final state: $Z\to \nu\bar \nu$ (0-leptons), $W\to \ell\nu$ (1-lepton), $Z\to \ell\ell$ (2-leptons). The 0-lepton channel shares the final state with the search presented in ref.~\cite{Aad:2015uga} (that we discussed above) and a small excess compatible with the one observed in that search has also been observed.  A similar search has been performed by CMS in ref.~\cite{Chatrchyan:2013zna}.

A search for $h\to b\bar b$ produced in association with dark matter has been presented in ref.~\cite{Aad:2015dva} where, unfortunately, the $bb$ invariant mass distribution is not shown.

Other interesting final states are $pp \to h + \slashed{E}_T \to (\mu\mu,\tau\tau,WW^*, ZZ^*) + \slashed{E}_T$. These modes are similar to the corresponding SM $pp \to h \to (\mu\mu,\tau\tau,WW^*, ZZ^*)$ ones, albeit with sizable extra missing energy that further reduces all backgrounds.

\section{Conclusions}
\label{sec:conclusions}

In two Higgs doublet model type II with vectorlike leptons, the decay modes of  heavy Higgs bosons can be dominated by cascade decays through the new leptons into $W$, $Z$ and Higgs bosons and SM leptons. These processes are listed in eqs.~(\ref{eq:Hn4}) and (\ref{eq:He4}) and corresponding Feynman diagrams are in figure~\ref{fig:topologies}.

After applying constraints from precision electroweak observables, searches for heavy Higgs bosons  and constraints on pair production of  vectorlike leptons obtained from searches for anomalous production of multilepton events we found that  branching ratios of  $H \to \nu_4 \nu_\mu$ and $H \to e_4 \mu$, where $e_4$ and $\nu_4$ are the  lightest new charged and neutral leptons can be as large as 50\%. These decay modes are especially relevant below the $t \bar t$ threshold and when the light Higgs boson ($h$) is SM-like so that $H \to ZZ,\; WW$ are suppressed or not present, competing only with $H\to b \bar b$  and for sufficiently heavy $H$ also with $H \to hh$.
 
 Furthermore, we found that each of the subsequent decay modes of $e_4$ and $\nu_4$: $e_4 \to W  \nu_\mu$, $e_4 \to Z \mu$,  $e_4 \to h \mu$ and $\nu_4 \to W \mu$, $\nu_4 \to Z \nu_\mu$, $\nu_4 \to h \nu_\mu$ can be close to 100\% providing many possible search opportunities. Among the most interesting  signatures are monojet, mono $Z$, mono Higgs, and $Z$ and Higgs bosons produced with a pair of charged leptons. Some of these signatures are almost background free. Combining this with potentially large production cross section for these processes presents great discover prospects at the LHC.

\acknowledgments 
The work of RD and EL was supported in part by the U.S. Department of Energy under grant number {DE}-SC0010120. RD was supported in part by the Ministry of Science, ICT and Planning (MSIP), South Korea, through the Brain Pool Program. RD also thanks the Galileo Galilei Institute for Theoretical Physics  for hospitality and support during part of this work.
 
\bibliographystyle{JHEP} 
\bibliography{vllsignals}

\providecommand{\href}[2]{#2}\begingroup\raggedright\begin{thebibliography}{10}

\bibitem{Dermisek:2015oja}
R.~Dermisek, E.~Lunghi and S.~Shin, \emph{{Two Higgs doublet model with
  vectorlike leptons and contributions to $pp\to WW$ and $H\to WW$}},
  \href{http://dx.doi.org/10.1007/JHEP02(2016)119}{\emph{JHEP} {\bf 02} (2016)
  119}, [\href{http://arxiv.org/abs/1509.04292}{{\tt 1509.04292}}].

\bibitem{Dermisek:2014qca}
R.~Dermisek, J.~P. Hall, E.~Lunghi and S.~Shin, \emph{{Limits on Vectorlike
  Leptons from Searches for Anomalous Production of Multi-Lepton Events}},
  \href{http://dx.doi.org/10.1007/JHEP12(2014)013}{\emph{JHEP} {\bf 12} (2014)
  013}, [\href{http://arxiv.org/abs/1408.3123}{{\tt 1408.3123}}].

\bibitem{Agashe:2014kda}
{\scshape Particle Data Group} collaboration, K.~A. Olive et~al., \emph{{Review
  of Particle Physics}},
  \href{http://dx.doi.org/10.1088/1674-1137/38/9/090001}{\emph{Chin. Phys.}
  {\bf C38} (2014) 090001}.

\bibitem{Heinemeyer:2013tqa}
{\scshape LHC Higgs Cross Section Working Group} collaboration, J.~R. Andersen
  et~al., \emph{{Handbook of LHC Higgs Cross Sections: 3. Higgs Properties}},
  \href{http://arxiv.org/abs/1307.1347}{{\tt 1307.1347}}.

\bibitem{Dermisek:2015vra}
R.~Dermisek, E.~Lunghi and S.~Shin, \emph{{Contributions of flavor violating
  couplings of a Higgs boson to $pp \to WW$}},
  \href{http://dx.doi.org/10.1007/JHEP08(2015)126}{\emph{JHEP} {\bf 08} (2015)
  126}, [\href{http://arxiv.org/abs/1503.08829}{{\tt 1503.08829}}].

\bibitem{Dermisek:2014cia}
R.~Dermisek, A.~Raval and S.~Shin, \emph{{Effects of vectorlike leptons on
  $h\to 4\ell$ and the connection to the muon g-2 anomaly}},
  \href{http://dx.doi.org/10.1103/PhysRevD.90.034023}{\emph{Phys. Rev.} {\bf
  D90} (2014) 034023}, [\href{http://arxiv.org/abs/1406.7018}{{\tt
  1406.7018}}].

\bibitem{Falkowski:2014ffa}
A.~Falkowski and R.~Vega-Morales, \emph{{Exotic Higgs decays in the golden
  channel}}, \href{http://dx.doi.org/10.1007/JHEP12(2014)037}{\emph{JHEP} {\bf
  12} (2014) 037}, [\href{http://arxiv.org/abs/1405.1095}{{\tt 1405.1095}}].

\bibitem{Kannike:2011ng}
K.~Kannike, M.~Raidal, D.~M. Straub and A.~Strumia, \emph{{Anthropic solution
  to the magnetic muon anomaly: the charged see-saw}},
  \href{http://dx.doi.org/10.1007/JHEP02(2012)106,
  10.1007/JHEP10(2012)136}{\emph{JHEP} {\bf 02} (2012) 106},
  [\href{http://arxiv.org/abs/1111.2551}{{\tt 1111.2551}}].

\bibitem{Dermisek:2013gta}
R.~Dermisek and A.~Raval, \emph{{Explanation of the Muon g-2 Anomaly with
  Vectorlike Leptons and its Implications for Higgs Decays}},
  \href{http://dx.doi.org/10.1103/PhysRevD.88.013017}{\emph{Phys. Rev.} {\bf
  D88} (2013) 013017}, [\href{http://arxiv.org/abs/1305.3522}{{\tt
  1305.3522}}].

\bibitem{Choudhury:2001hs}
D.~Choudhury, T.~M.~P. Tait and C.~E.~M. Wagner, \emph{{Beautiful mirrors and
  precision electroweak data}},
  \href{http://dx.doi.org/10.1103/PhysRevD.65.053002}{\emph{Phys. Rev.} {\bf
  D65} (2002) 053002}, [\href{http://arxiv.org/abs/hep-ph/0109097}{{\tt
  hep-ph/0109097}}].

\bibitem{Dermisek:2011xu}
R.~Dermisek, S.-G. Kim and A.~Raval, \emph{{New Vector Boson Near the Z-pole
  and the Puzzle in Precision Electroweak Data}},
  \href{http://dx.doi.org/10.1103/PhysRevD.84.035006}{\emph{Phys. Rev.} {\bf
  D84} (2011) 035006}, [\href{http://arxiv.org/abs/1105.0773}{{\tt
  1105.0773}}].

\bibitem{Dermisek:2012qx}
R.~Dermisek, S.-G. Kim and A.~Raval, \emph{{Z' near the Z-pole}},
  \href{http://dx.doi.org/10.1103/PhysRevD.85.075022}{\emph{Phys. Rev.} {\bf
  D85} (2012) 075022}, [\href{http://arxiv.org/abs/1201.0315}{{\tt
  1201.0315}}].

\bibitem{Batell:2012ca}
B.~Batell, S.~Gori and L.-T. Wang, \emph{{Higgs Couplings and Precision
  Electroweak Data}},
  \href{http://dx.doi.org/10.1007/JHEP01(2013)139}{\emph{JHEP} {\bf 01} (2013)
  139}, [\href{http://arxiv.org/abs/1209.6382}{{\tt 1209.6382}}].

\bibitem{Dermisek:2012as}
R.~Dermisek, \emph{{Insensitive Unification of Gauge Couplings}},
  \href{http://dx.doi.org/10.1016/j.physletb.2012.06.037}{\emph{Phys. Lett.}
  {\bf B713} (2012) 469--472}, [\href{http://arxiv.org/abs/1204.6533}{{\tt
  1204.6533}}].

\bibitem{Dermisek:2012ke}
R.~Dermisek, \emph{{Unification of gauge couplings in the standard model with
  extra vectorlike families}},
  \href{http://dx.doi.org/10.1103/PhysRevD.87.055008}{\emph{Phys. Rev.} {\bf
  D87} (2013) 055008}, [\href{http://arxiv.org/abs/1212.3035}{{\tt
  1212.3035}}].

\bibitem{Babu:1996zv}
K.~S. Babu and J.~C. Pati, \emph{{The Problems of unification mismatch and low
  alpha**3: A Solution with light vector - like matter}},
  \href{http://dx.doi.org/10.1016/0370-2693(96)00809-X}{\emph{Phys. Lett.} {\bf
  B384} (1996) 140--150}, [\href{http://arxiv.org/abs/hep-ph/9606215}{{\tt
  hep-ph/9606215}}].

\bibitem{BasteroGil:1999dx}
M.~Bastero-Gil and B.~Brahmachari, \emph{{Semiperturbative unification with
  extra vector - like families}},
  \href{http://dx.doi.org/10.1016/S0550-3213(00)00069-9}{\emph{Nucl. Phys.}
  {\bf B575} (2000) 35--60}, [\href{http://arxiv.org/abs/hep-ph/9907318}{{\tt
  hep-ph/9907318}}].

\bibitem{Kolda:1996ea}
C.~F. Kolda and J.~March-Russell, \emph{{Low-energy signatures of
  semiperturbative unification}},
  \href{http://dx.doi.org/10.1103/PhysRevD.55.4252}{\emph{Phys. Rev.} {\bf D55}
  (1997) 4252--4261}, [\href{http://arxiv.org/abs/hep-ph/9609480}{{\tt
  hep-ph/9609480}}].

\bibitem{Barr:2012ma}
S.~M. Barr and H.-Y. Chen, \emph{{A Simple Grand Unified Relation between
  Neutrino Mixing and Quark Mixing}},
  \href{http://dx.doi.org/10.1007/JHEP11(2012)092}{\emph{JHEP} {\bf 11} (2012)
  092}, [\href{http://arxiv.org/abs/1208.6546}{{\tt 1208.6546}}].

\bibitem{Martin:2009bg}
S.~P. Martin, \emph{{Extra vector-like matter and the lightest Higgs scalar
  boson mass in low-energy supersymmetry}},
  \href{http://dx.doi.org/10.1103/PhysRevD.81.035004}{\emph{Phys. Rev.} {\bf
  D81} (2010) 035004}, [\href{http://arxiv.org/abs/0910.2732}{{\tt
  0910.2732}}].

\bibitem{Bae:2012ir}
K.~J. Bae, T.~H. Jung and H.~D. Kim, \emph{{125 GeV Higgs boson as a
  pseudo-Goldstone boson in supersymmetry with vectorlike matters}},
  \href{http://dx.doi.org/10.1103/PhysRevD.87.015014}{\emph{Phys. Rev.} {\bf
  D87} (2013) 015014}, [\href{http://arxiv.org/abs/1208.3748}{{\tt
  1208.3748}}].

\bibitem{Ellis:2014dza}
S.~A.~R. Ellis, R.~M. Godbole, S.~Gopalakrishna and J.~D. Wells, \emph{{Survey
  of vector-like fermion extensions of the Standard Model and their
  phenomenological implications}},
  \href{http://dx.doi.org/10.1007/JHEP09(2014)130}{\emph{JHEP} {\bf 09} (2014)
  130}, [\href{http://arxiv.org/abs/1404.4398}{{\tt 1404.4398}}].

\bibitem{Kumar:2015tna}
N.~Kumar and S.~P. Martin, \emph{{Vectorlike leptons at the Large Hadron
  Collider}}, \href{http://dx.doi.org/10.1103/PhysRevD.92.115018}{\emph{Phys.
  Rev.} {\bf D92} (2015) 115018}, [\href{http://arxiv.org/abs/1510.03456}{{\tt
  1510.03456}}].

\bibitem{CMS:bxa}
CMS{\ }Collaboration, \emph{{Update on the search for the standard model Higgs
  boson in pp collisions at the LHC decaying to W + W in the fully leptonic
  final state}},  \href{http://arxiv.org/abs/CMS-PAS-HIG-13-003}{{\tt
  CMS-PAS-HIG-13-003}}.

\bibitem{Chatrchyan:2013iaa}
{\scshape CMS} collaboration, S.~Chatrchyan et~al., \emph{{Measurement of Higgs
  boson production and properties in the WW decay channel with leptonic final
  states}}, \href{http://dx.doi.org/10.1007/JHEP01(2014)096}{\emph{JHEP} {\bf
  01} (2014) 096}, [\href{http://arxiv.org/abs/1312.1129}{{\tt 1312.1129}}].

\bibitem{CMS:2014onr}
CMS{\ }Collaboration, \emph{{Search for an Higgs Like resonance in the diphoton
  mass spectra above 150 GeV with 8 TeV data}},
  \href{http://arxiv.org/abs/CMS-PAS-HIG-14-006}{{\tt CMS-PAS-HIG-14-006}}.

\bibitem{Dermisek:2013cxa}
R.~Dermisek, J.~P. Hall, E.~Lunghi and S.~Shin, \emph{{A New Avenue to Charged
  Higgs Discovery in Multi-Higgs Models}},
  \href{http://dx.doi.org/10.1007/JHEP04(2014)140}{\emph{JHEP} {\bf 04} (2014)
  140}, [\href{http://arxiv.org/abs/1311.7208}{{\tt 1311.7208}}].

\bibitem{Song:2014lua}
J.~Song and Y.~W. Yoon, \emph{{Gigantic diphoton rate of heavy Higgs bosons in
  the aligned two Higgs doublet models with small tanβ}},
  \href{http://dx.doi.org/10.1103/PhysRevD.91.113012}{\emph{Phys. Rev.} {\bf
  D91} (2015) 113012}, [\href{http://arxiv.org/abs/1412.5610}{{\tt
  1412.5610}}].

\bibitem{Khachatryan:2014ira}
{\scshape CMS} collaboration, V.~Khachatryan et~al., \emph{{Observation of the
  diphoton decay of the Higgs boson and measurement of its properties}},
  \href{http://dx.doi.org/10.1140/epjc/s10052-014-3076-z}{\emph{Eur. Phys. J.}
  {\bf C74} (2014) 3076}, [\href{http://arxiv.org/abs/1407.0558}{{\tt
  1407.0558}}].

\bibitem{ATLAS:2011ae}
{\scshape ATLAS} collaboration, G.~Aad et~al., \emph{{Search for the Higgs
  boson in the $H -> WW -> lvjj$ decay channel in pp collisions at sqrt{s} = 7
  TeV with the ATLAS detector}},
  \href{http://dx.doi.org/10.1103/PhysRevLett.107.231801}{\emph{Phys. Rev.
  Lett.} {\bf 107} (2011) 231801}, [\href{http://arxiv.org/abs/1109.3615}{{\tt
  1109.3615}}].

\bibitem{Chatrchyan:2013yoa}
{\scshape CMS} collaboration, S.~Chatrchyan et~al., \emph{{Search for a
  standard-model-like Higgs boson with a mass in the range 145 to 1000 GeV at
  the LHC}}, \href{http://dx.doi.org/10.1140/epjc/s10052-013-2469-8}{\emph{Eur.
  Phys. J.} {\bf C73} (2013) 2469}, [\href{http://arxiv.org/abs/1304.0213}{{\tt
  1304.0213}}].

\bibitem{CMS:2015lda}
CMS{\ }Collaboration, \emph{{Search for a Standard Model-like Higgs Boson
  Decaying into $WW \to l \nu q\bar{q}'$ in Exclusive Jet Bins in pp Collisions
  at $\sqrt{s}$ = 8 TeV}},  \href{http://arxiv.org/abs/CMS-PAS-HIG-14-008}{{\tt
  CMS-PAS-HIG-14-008}}.

\bibitem{ATLAS_WW8}
ATLAS{\ }Collaboration, \emph{{Search for $WW$, $WZ$, and $ZZ$ resonances in
  $pp$ collisions at $\sqrt{s} = 8$ TeV with the ATLAS detector}},
  \href{http://arxiv.org/abs/ATLAS-CONF-2015-045}{{\tt ATLAS-CONF-2015-045}}.

\bibitem{ATLAS_WW13}
ATLAS{\ }Collaboration, \emph{{Search for $WW/WZ$ resonance production in the
  $\ell\nu qq$ final state at $\sqrt{s}=13\,$ TeV with the ATLAS detector at
  the LHC}},  \href{http://arxiv.org/abs/ATLAS-CONF-2015-075}{{\tt
  ATLAS-CONF-2015-075}}.

\bibitem{Aad:2015zva}
{\scshape ATLAS} collaboration, G.~Aad et~al., \emph{{Search for new phenomena
  in final states with an energetic jet and large missing transverse momentum
  in pp collisions at $\sqrt{s}=$8 TeV with the ATLAS detector}},
  \href{http://dx.doi.org/10.1140/epjc/s10052-015-3517-3,
  10.1140/epjc/s10052-015-3639-7}{\emph{Eur. Phys. J.} {\bf C75} (2015) 299},
  [\href{http://arxiv.org/abs/1502.01518}{{\tt 1502.01518}}].

\bibitem{Khachatryan:2014rra}
{\scshape CMS} collaboration, V.~Khachatryan et~al., \emph{{Search for dark
  matter, extra dimensions, and unparticles in monojet events in
  proton–proton collisions at $\sqrt{s} = 8$ TeV}},
  \href{http://dx.doi.org/10.1140/epjc/s10052-015-3451-4}{\emph{Eur. Phys. J.}
  {\bf C75} (2015) 235}, [\href{http://arxiv.org/abs/1408.3583}{{\tt
  1408.3583}}].

\bibitem{Khachatryan:2015pba}
{\scshape CMS} collaboration, V.~Khachatryan et~al., \emph{{Measurements of the
  $\mathrm{Z}$ $\mathrm{Z}$ production cross sections in the $2\mathrm{l} 2\nu
  $ channel in proton–proton collisions at $\sqrt{s} = 7$ and $8~\mathrm{TeV}
  $ and combined constraints on triple gauge couplings}},
  \href{http://dx.doi.org/10.1140/epjc/s10052-015-3706-0}{\emph{Eur. Phys. J.}
  {\bf C75} (2015) 511}, [\href{http://arxiv.org/abs/1503.05467}{{\tt
  1503.05467}}].

\bibitem{Aad:2015kna}
{\scshape ATLAS} collaboration, G.~Aad et~al., \emph{{Search for an additional,
  heavy Higgs boson in the $H\rightarrow ZZ$ decay channel at $\sqrt{s}$ = 8
  TeV in $pp$ collision data with the ATLAS detector}},
  \href{http://arxiv.org/abs/1507.05930}{{\tt 1507.05930}}.

\bibitem{Aad:2014vka}
{\scshape ATLAS} collaboration, G.~Aad et~al., \emph{{Search for dark matter in
  events with a Z boson and missing transverse momentum in pp collisions at
  $\sqrt{s}$=8 TeV with the ATLAS detector}},
  \href{http://dx.doi.org/10.1103/PhysRevD.90.012004}{\emph{Phys. Rev.} {\bf
  D90} (2014) 012004}, [\href{http://arxiv.org/abs/1404.0051}{{\tt
  1404.0051}}].

\bibitem{CMS:2015kha}
CMS{\ }Collaboration, \emph{{Search for dark matter and unparticles produced in
  association with a Z boson in pp collisions at sqrt(s) = 8 TeV}},
  \href{http://arxiv.org/abs/CMS-PAS-EXO-12-054}{{\tt CMS-PAS-EXO-12-054}}.

\bibitem{Aad:2014iia}
{\scshape ATLAS} collaboration, G.~Aad et~al., \emph{{Search for Invisible
  Decays of a Higgs Boson Produced in Association with a Z Boson in ATLAS}},
  \href{http://dx.doi.org/10.1103/PhysRevLett.112.201802}{\emph{Phys.Rev.Lett.}
  {\bf 112} (2014) 201802}, [\href{http://arxiv.org/abs/1402.3244}{{\tt
  1402.3244}}].

\bibitem{Aad:2015dha}
{\scshape ATLAS} collaboration, G.~Aad et~al., \emph{{Search for heavy lepton
  resonances decaying to a $Z$ boson and a lepton in $pp$ collisions at
  $\sqrt{s}=8$ TeV with the ATLAS detector}},
  \href{http://dx.doi.org/10.1007/JHEP09(2015)108}{\emph{JHEP} {\bf 09} (2015)
  108}, [\href{http://arxiv.org/abs/1506.01291}{{\tt 1506.01291}}].

\bibitem{Khachatryan:2015cwa}
{\scshape CMS} collaboration, V.~Khachatryan et~al., \emph{{Search for a Higgs
  Boson in the Mass Range from 145 to 1000 GeV Decaying to a Pair of W or Z
  Bosons}}, \href{http://dx.doi.org/10.1007/JHEP10(2015)144}{\emph{JHEP} {\bf
  10} (2015) 144}, [\href{http://arxiv.org/abs/1504.00936}{{\tt 1504.00936}}].

\bibitem{CMS:2014xja}
{\scshape CMS} collaboration, V.~Khachatryan et~al., \emph{{Measurement of the
  $pp \to ZZ$ production cross section and constraints on anomalous triple
  gauge couplings in four-lepton final states at $\sqrt s=$8 TeV}},
  \href{http://dx.doi.org/10.1016/j.physletb.2014.11.059}{\emph{Phys. Lett.}
  {\bf B740} (2015) 250--272}, [\href{http://arxiv.org/abs/1406.0113}{{\tt
  1406.0113}}].

\bibitem{Aad:2015yga}
{\scshape ATLAS} collaboration, G.~Aad et~al., \emph{{Search for Dark Matter in
  Events with Missing Transverse Momentum and a Higgs Boson Decaying to Two
  Photons in $pp$ Collisions at $\sqrt{s}=8$ TeV with the ATLAS Detector}},
  \href{http://dx.doi.org/10.1103/PhysRevLett.115.131801}{\emph{Phys. Rev.
  Lett.} {\bf 115} (2015) 131801}, [\href{http://arxiv.org/abs/1506.01081}{{\tt
  1506.01081}}].

\bibitem{Aad:2015uga}
{\scshape ATLAS} collaboration, G.~Aad et~al., \emph{{Search for invisible
  decays of the Higgs boson produced in association with a hadronically
  decaying vector boson in $pp$ collisions at $\sqrt{s}$ = 8 TeV with the ATLAS
  detector}},
  \href{http://dx.doi.org/10.1140/epjc/s10052-015-3551-1}{\emph{Eur. Phys. J.}
  {\bf C75} (2015) 337}, [\href{http://arxiv.org/abs/1504.04324}{{\tt
  1504.04324}}].

\bibitem{Chatrchyan:2014tja}
{\scshape CMS} collaboration, S.~Chatrchyan et~al., \emph{{Search for invisible
  decays of Higgs bosons in the vector boson fusion and associated ZH
  production modes}},
  \href{http://dx.doi.org/10.1140/epjc/s10052-014-2980-6}{\emph{Eur. Phys. J.}
  {\bf C74} (2014) 2980}, [\href{http://arxiv.org/abs/1404.1344}{{\tt
  1404.1344}}].

\bibitem{Aad:2014xzb}
{\scshape ATLAS} collaboration, G.~Aad et~al., \emph{{Search for the $b\bar{b}$
  decay of the Standard Model Higgs boson in associated $(W/Z)H$ production
  with the ATLAS detector}},
  \href{http://dx.doi.org/10.1007/JHEP01(2015)069}{\emph{JHEP} {\bf 01} (2015)
  069}, [\href{http://arxiv.org/abs/1409.6212}{{\tt 1409.6212}}].

\bibitem{Chatrchyan:2013zna}
{\scshape CMS} collaboration, S.~Chatrchyan et~al., \emph{{Search for the
  standard model Higgs boson produced in association with a W or a Z boson and
  decaying to bottom quarks}},
  \href{http://dx.doi.org/10.1103/PhysRevD.89.012003}{\emph{Phys. Rev.} {\bf
  D89} (2014) 012003}, [\href{http://arxiv.org/abs/1310.3687}{{\tt
  1310.3687}}].

\bibitem{Aad:2015dva}
{\scshape ATLAS} collaboration, G.~Aad et~al., \emph{{Search for dark matter
  produced in association with a Higgs boson decaying to two bottom quarks in
  $pp$ collisions at $\sqrt{s} = 8$ TeV with the ATLAS detector}},
  \href{http://arxiv.org/abs/1510.06218}{{\tt 1510.06218}}.

\end{thebibliography}\endgroup

\end{document}